\documentclass[11pt]{revtex4}
\usepackage{amssymb}
\usepackage{latexsym}
\usepackage{epsfig}

\begin{document}

\title{Entropic Corrections to Einstein Equations}
\author{S. H. Hendi$^{1,2}$\footnote{hendi@mail.yu.ac.ir} and A. Sheykhi$^{2,3}
$\footnote{sheykhi@mail.uk.ac.ir}} \affiliation{$^1$ Physics
Department, College of Sciences, Yasouj University, Yasouj
75914, Iran\\
$^2$ Research Institute for Astronomy and Astrophysics of Maragha (RIAAM),
Maragha, Iran\\
$^3$Department of Physics, Shahid Bahonar University, P.O. Box 76175-132,
Kerman, Iran}

\begin{abstract}
Considering the general quantum corrections to the area law of
black hole entropy and adopting the viewpoint that gravity
interprets as an entropic force, we derive the modified forms of
MOND theory of gravitation and Einstein field equations. As two
special cases we study the logarithmic and power-law corrections
to entropy and find the explicit form of the obtained modified
equations.

\end{abstract}

\maketitle

%%%%%%%%%%%%%%%%%%%%%%%%%%%%%%%%%%%%%%%%%%%%%%%%%%%%%%%%%%%%%%%%

\section{Introduction}

The discovery of the black holes thermodynamics in 1970's, implies that
there should be some deep connection between the laws of thermodynamics and
gravity. According to the black hole thermodynamics a black hole has an
entropy proportional to its horizon area and a temperature proportional to
its surface gravity, and the entropy and temperature together with the mass
of the black hole satisfy the first law of thermodynamics \cite{HB}.
Jacobson was the first who revealed the deep connection between
thermodynamics and gravity by deriving the Einstein field equation from the
first law of thermodynamics \cite{Jac}. Following Jacobson, however, several
recent investigations have shown that there is indeed a deeper connection
between gravitational dynamics and horizon thermodynamics. It has been shown
that the gravitational field equations in a wide variety of theories, when
evaluated on a horizon, reduce to the first law of thermodynamics and vice
versa. This result, first pointed out in \cite{Pad1}, has now been
demonstrated in various theory including f(R) gravity \cite{Elin},
cosmological setups \cite{Cai2,Cai3,CaiKim,Wang,Cai33,Shey0}, and in
braneworld scenarios \cite{Shey1,Shey2}. For a recent review on the
thermodynamical aspects of gravity and complete list of references see \cite%
{Pad2}. The deep connection between horizon thermodynamics and gravitational
dynamics, help us to understand why the field equations should encode
information about horizon thermodynamics. These results prompt people to
take a statistical physics point of view on gravity.

An interesting new proposal on the statistical mechanical origin
of the thermodynamic nature of gravity was recently suggested by
Verlinde \cite{Ver} who interpreted gravity as an entropic force
caused by the changes in the information associated with the
positions of material bodies. It is notable that, difficulty of
unification of gravity with quantum mechanics is one of the basic
motivations that leads Verlinde to propose this new pattern.
Verlinde made several interesting observations. First, with the
assumption of the entropic force together with the Unruh
temperature, he derived the second law of Newton. Second, by
taking into account the entropic force together with the
holographic principle and the equipartition law of energy he
obtained the Newton's law of gravitation. A relativistic
generalization of this argument directly leads to the Einstein
equations. Similar discoveries about statistical origin of gravity
and combination of the equipartition law for horizon degrees of
freedom with the Smarr formula \cite{Smarr} were also studied in
\cite{TPad3,Banerjee1}. This may imply that the entropy is to link
general relativity with the statistical description of unknown
spacetime microscopic structure when the horizon is present.

Inspired by Bekenstein's entropy bound, Verlinde proposed that
when a test particle with mass $m$ approaches a holographic screen
from a distance $\triangle x$, the change of entropy on the
holographic screen is
\begin{equation}  \label{s0}
\triangle S=2\pi k_B \frac{mc}{\hbar} \triangle x.
\end{equation}
The entropic force can arise in the direction of increasing entropy and is
proportional to the temperature, $F=T \triangle S/ \triangle x$. Verlinde's
derivation of Newton's law of gravitation at the very least offers a strong
analogy with a well understood statistical mechanism. Therefore, this
derivation opens a new window to understand gravity from the first
principles. The study on the entropic force has raised a lot of attention
recently (see \cite{Cai4,Other, sheyECFE,Ling,Modesto,Yi, HS} and references
therein).

According to Verlinde's discussion, the number of bits on holographic screen
can be specified as $N=4S=\frac{Ac^{3}}{G\hbar }$. Indeed, the derivation of
Newton's law of gravity as well as Einstein equations depend on the
entropy-area relationship $S=A/4\ell _{p}^{2}$, where $A=4\pi R^{2}$
represents the area of the horizon and $\ell _{p}^{2}=G\hbar /c^{3}$ is the
Planck length. However, the entropy-area relation can be modified from the
inclusion of quantum effects. Consider the general corrections to area law,
we write down the general entropy-corrected relation as
\begin{equation}
S=\frac{A}{4\ell _{p}^{2}}+{s}(A),  \label{S2}
\end{equation}
where $s(A)$ stands the general correction terms. Throughout this
paper we set $k_{B}=1$ for simplicity. Two well-known quantum
corrections to the area law have been introduced in the
literatures, namely, logarithmic and power-law corrections.
Logarithmic corrections, arises from loop quantum gravity due to
thermal equilibrium fluctuations and quantum fluctuations
\cite{Meis,Zhang,LogCorrection0,LogCorrection1,LogCorrection21,LogCorrection22,LogCorrection23,LogCorrection31,LogCorrection32,LogCorrection33,LogCorrection34,LogCorrection35},
\begin{equation}
S=\frac{A}{4\ell _{p}^{2}}-\beta \ln {\frac{A}{4\ell
_{p}^{2}}}+\gamma \frac{\ell _{p}^{2}}{A}+\mathrm{const},
\label{S1}
\end{equation}
where $\beta $ and $\gamma $ are unknown dimensionless constants.
The issue of the value of $\beta$ and $\gamma$ is highly
disputatious and one can find different interpretations in the
literature. By using trace anomaly, tunneling method, path
integral or other quantum and semiclassical type approaches, many
authors has been discussed about the value of the parameter
$\beta$ for the Schwarzschild black hole (see
\cite{LogCorrection21,LogCorrection22,LogCorrection23} for more
details). In addition, one can find different values for $\beta$
and $\gamma$ in literature
\cite{LogCorrection31,LogCorrection32,LogCorrection33,LogCorrection34,LogCorrection35}.
Regardless of different reported value for parameters $\beta$ and
$\gamma$, in this paper we can derive the modified gravitational
field equations. One can find that these distinctions do not
affect on our discussion due to our general analysis.

Another form of correction to area law, namely the power-law
correction, appears in dealing with the entanglement of quantum
fields in and out the horizon. The entanglement entropy of the
ground state obeys the Hawking area law. Only the excited state
contributes to the correction, and more excitations produce more
deviation from the area law \cite{sau1} (also see \cite{sau2} for
a review on the origin of black hole entropy through
entanglement). The power-law corrected entropy is written as
\cite{Sau,pavon1}
\begin{equation}
S=\frac{A}{4\ell _{p}^{2}}\left[ 1-K_{\alpha }A^{1-\alpha /2}\right] ,
\label{plec}
\end{equation}
where $\alpha $ is a dimensionless constant whose value ranges as $2<\alpha
<4$ \cite{Sau}. Here
\begin{equation}
K_{\alpha }=\frac{\alpha (4\pi )^{\alpha /2-1}}{(4-\alpha
)r_{c}^{2-\alpha }} ,  \label{kalpha}
\end{equation}
where $r_{c}$ is the crossover scale. The second term in Eq. (\ref{plec})
can be regarded as a power-law correction to the area law, resulting from
entanglement, when the wave-function of the field is chosen to be a
superposition of ground state and exited state \cite{Sau}.

\section{Entropic corrections to MOND theory}

Modified Newtonian dynamics (briefly abbreviated as MOND) is a hypothesis
that proposes a modification of Newton's law of gravity to explain the
galaxy rotation problem. When the uniform velocity of rotation of galaxies
was first observed, it was unexpected because Newtonian theory predicts that
objects that are further out will have lower velocities.

In 1983, M. Milgrom \cite{Milgrom} suggested the MOND theory which
appears to be highly successful for explaining the observed
anomalous rotational-velocity . But unfortunately MOND theory
lacks theoretical support. In fact, the MOND theory is (empirical)
modification of Newtonian dynamics through modification in the
kinematical acceleration term `$a$' (which is normally taken as
$a=v^{2}/r$) as effective kinematic acceleration
$a_{\mathrm{eff}}=a \mu (\frac{a}{a_{0}})$ , wherein the
$\mu$-function is identical to one for usual-values of
accelerations but equals to $\frac{a}{a_{0}}$($\ll 1$) if the
acceleration `$a$' is extremely low, lower than a critical value
$a_{0}$ ($10^{-10}$ $m/s^{2}$). At large distance at the galaxy
out skirt, the kinematical acceleration `$a$' is extremely small,
smaller than $10^{-10}$ $m/s^{2}$ , i.e., $a\ll a_{0}$, hence the
function $\mu (\frac{a}{a_{0}})=\frac{a}{a_{0}}$. Consequently,
the velocity of star on circular orbit from the galaxy-center is
constant and does not depend on the distance; the rotational-curve
is flat, as it observed.

In this section, by applying the modified entropy-area relation (\ref{S2}),
we reproduce the corrected MOND theory following the methods of \cite{Neto}.
We adopt the assumption that MOND theory can be viewed as an entropic force.
We suppose that below a critical temperature, the cooling of the holographic
screen is not homogeneous. Also we assume that the fraction of bits with
zero energy is given by \cite{Neto}
\begin{equation}
\frac{N_{0}}{N}=1-\frac{T}{T_{c}},  \label{Mo1}
\end{equation}
where it is a second order phase transitions theory and usual
relation of the critical phenomena. Since we consider $T_{c}$ is a
critical temperature in which, below this, the zero energy
phenomenon for some bits starts to occur, we keep $N_{0}=0$ for
$T\geq T_{c}$. In other words for temperature $T<T_{c}$, the
number of bits with nonzero energy is
\begin{equation}
N-N_{0}=N\frac{T}{T_{c}},  \label{Mo2}
\end{equation}
and the generalized equipartition law of energy is
\begin{equation}
E=\frac{1}{2}(N-N_{0})T.  \label{GEqui}
\end{equation}
Substituting Eq. (\ref{Mo2}) in the generalized equipartition law of energy,
we obtain
\begin{equation}
E=\frac{1}{2}N\frac{T}{T_{c}}T.  \label{GEqui2}
\end{equation}
One can combine Eq. (\ref{GEqui2}) with $E=Mc^{2}$ to obtain temperature
\begin{equation}
T^{2}=\frac{2Mc^{2}T_{c}}{N}.  \label{T2}
\end{equation}
Using the Unruh temperature formula, $T=\frac{1}{2\pi }\frac{\hbar a}{c}$,
as well as Eq. (\ref{T2}), we get
\begin{equation}
a^{2}=\frac{8\pi ^{2}c^{2}}{\hbar ^{2}}\frac{Mc^{2}T_{c}}{N}.  \label{a2}
\end{equation}
According to statistical mechanics the entropy of a system is
proportional to the number of bits. Therefore, considering the
general entropy-corrected relation (\ref{S2}), we can write the
relation between $A$ and $N$ as
\begin{equation}
N=4S=\frac{1}{\ell _{p}^{2}}\left[ A+4\ell _{p}^{2}s(A)\right] .  \label{NA}
\end{equation}
Here, we use Eq. (\ref{NA}) and the fact that $A=4\pi R^{2}$, to rewrite Eq.
(\ref{a2}) in the following form
\begin{equation}
a^{2}\left( \frac{4\pi R^{2}}{\ell _{p}^{2}}\right) \left[
1+\frac{\ell _{p}^{2}}{\pi R^{2}}s(A)\right] =\frac{8\pi
^{2}c^{2}}{\hbar ^{2}} Mc^{2}T_{c}.  \label{ModMOND1}
\end{equation}
After some algebraic manipulations and using $(1+\mathcal{\alpha}
)^{n}\approx 1+n\mathcal{\alpha }$ for $\mathcal{\alpha }\ll 1$
(where $\mathcal{\alpha }=\frac{\ell _{p}^{2}}{\pi R^{2}}s(A)$ and
$n=-1$), we derive the entropic corrections to MOND theory as
\begin{equation}
a\left( \frac{a}{\frac{2\pi cT_{c}}{\hbar }}\right) =\frac{GM}{R^{2}}\left[
1-\frac{\ell _{p}^{2}}{\pi R^{2}}s(A)\right] ,  \label{ModMOND2}
\end{equation}
or equivalently
\begin{equation}
a\left( \frac{a}{a_{0}}\right) =\frac{GM}{R^{2}}\left[ 1-\frac{\ell _{p}^{2}%
}{\pi R^{2}}s(A)\right] ,  \label{ModMOND}
\end{equation}
where $a_{0}=2\pi c\hbar ^{-1}T_{c}$. Now, we want to calculate the explicit
form of quantum corrections to this formula. In order to do this we need to
use Eqs. (\ref{S1}) and (\ref{plec}), which lead to the following relations
\begin{equation}
a\left( \frac{a}{a_{0}}\right) =\frac{GM}{R^{2}}\left[ 1+\frac{\beta \ell
_{p}^{2}}{\pi R^{2}}\ln {\frac{\pi R^{2}}{\ell _{p}^{2}}}-\frac{\gamma \ell
_{p}^{4}}{4\pi ^{2}R^{4}}\right] ,  \label{ModMOND3}
\end{equation}
\begin{equation}
a\left( \frac{a}{a_{0}}\right) =\frac{GM}{R^{2}}\left[
1+\frac{\alpha }{(4-\alpha )}\left( \frac{r_{c}}{R}\right)
^{\alpha -2}\right] . \label{ModMOND33}
\end{equation}
Eqs. (\ref{ModMOND3}) and (\ref{ModMOND33}) are the corrected MOND theory
corresponding to the logarithmic and the power-law entropy corrections,
respectively.

\section{Entropic corrections to Einstein Equations}

Finally, we reach to our main task in this paper, namely
considering the general entropic corrections to Einstein field
equations. In particular, we investigate the influence of the
number of bits on the holographic screen on the Einstein field
equation. In general, the entropy is proportional to the number of
bits living on the holographic screen. Therefore, we propose the
number of bits-entropy relation has the following form%
\begin{equation}
N=4S.  \label{NS}
\end{equation}
Consider the general entropy-corrected relation (\ref{S2}), and
using Eq. (\ref{NA}), we can write the relation between $A$ and
$N$ in differential form
\begin{equation}
dN=\frac{1}{\ell _{p}^{2}}\left[ 1+4\ell _{p}^{2}\frac{\partial
s}{\partial A}\right] dA.  \label{dNdA}
\end{equation}
Employing the more general equipartition law we have \cite{Pad4}
\begin{equation}
M=\frac{1}{2}\int TdN,  \label{Equipart}
\end{equation}
where $T=\frac{\hbar }{2\pi c^{3}}e^{\phi }N^{b}\triangledown _{b}\phi $
(see \cite{Ver} for more details). Substituting $T$ and $dN$, we obtain
\begin{equation}
M=\frac{\hbar }{4\pi c^{3}\ell _{p}^{2}}\int e^{\phi }\left[
1+4\ell _{p}^{2} \frac{\partial s}{\partial A}\right]
\triangledown \phi .dA.  \label{M1}
\end{equation}
Using the relation between $\phi $ and killing vector $\xi ^{a}$
\cite{Waldbook} as well as the Stokes theorem, and following the
same logic of \cite{Ver}, we can obtain
\begin{equation}
\int e^{\phi }\triangledown \phi .dA=\int R_{\mu \nu }N^{\mu }\xi ^{\nu }dV,
\label{Phikilling}
\end{equation}
or equivalently
\begin{equation}
M=\frac{\hbar }{4\pi c^{3}\ell _{p}^{2}}\left[ \int R_{\mu \nu }N^{\mu }\xi
^{\nu }dV+4\ell _{p}^{2}\int e^{\phi }\left( \frac{\partial s}{\partial A}%
\right) \triangledown \phi .dA\right] .  \label{M2}
\end{equation}
On the other hand, $M$ can be expressed as an integral over the enclosed
volume of certain components of stress energy tensor $T_{ab}$
\begin{equation}
M=2\int \left( T_{\mu \nu }-\frac{1}{2}Tg_{\mu \nu }\right) N^{\mu }\xi
^{\nu }dV,  \label{M22}
\end{equation}
Equating Eqs. (\ref{M2}) and (\ref{M22}), with $c^{3}\ell
_{p}^{2}/\hbar =G$, we find
\begin{equation}
2\int \left( T_{\mu \nu }-\frac{1}{2}Tg_{\mu \nu }\right) N^{\mu }\xi ^{\nu
}dV=\frac{1}{4\pi G}\left[ \int R_{\mu \nu }N^{\mu }\xi ^{\nu }dV+4\ell
_{p}^{2}\int e^{\phi }\left( \frac{\partial s}{\partial A}\right)
\triangledown \phi .dA\right].  \label{ModEE}
\end{equation}
One can rewrite Eq. (\ref{ModEE}) in the following form
\begin{equation}
\int \left[ R_{\mu \nu }-8\pi G\left( T_{\mu \nu }-\frac{1}{2}Tg_{\mu \nu
}\right) \right] N^{\mu }\xi ^{\nu }dV=-4\ell _{p}^{2}\int e^{\phi }\left(
\frac{\partial s}{\partial A}\right) \triangledown \phi .dA  \label{ModE}
\end{equation}
This is the entropic correction Einstein equations. The
right-hand-side of Eq. (\ref{ModE}) is an additional term compared
with Verlinde's result \cite{Ver}. This term is caused by the
correction to the number of bits on the holographic screen (or
correction to area law) which brings a surface correction to the
Einstein field equations. In order to simplify the
integral of right-hand-side, the functional form of the correction terms $%
s(A)$ as well as the value of $e^{\phi }\triangledown \phi $ should be
specified. For example, when we consider power-law correction to entropy,
the function $s(A)$ is
\begin{equation}
s(A)=-\frac{1}{4\ell _{p}^{2}}K_{\alpha }A^{2-\alpha /2}.  \label{SPL}
\end{equation}
Taking the power-law correction to entropy and assuming that the spacetime
is vacuum static spherically symmetric, in which describes with the
Schwarzschild metric, we reach
\begin{equation}
\int e^{\phi }\left( \frac{\partial s}{\partial A}\right) \triangledown \phi
.dA=-(\frac{\alpha -4}{\alpha -2})\frac{K_{\alpha }GM}{4\ell _{p}^{2}}%
A^{1-\alpha /2},  \label{SPL2}
\end{equation}
Inserting this into Eq. (\ref{ModE}) and using (\ref{M22}) we find
\begin{equation}
\int \left[ R_{\mu \nu }-8\pi G\left( T_{\mu \nu }-\frac{1}{2}Tg_{\mu \nu
}\right) -2\left( \frac{\alpha -4}{\alpha -2}\right) \frac{K_{\alpha }G}{%
A^{\alpha /2-1}}\left( T_{\mu \nu }-\frac{1}{2}Tg_{\mu \nu }\right) \right]
N^{\mu }\xi ^{\nu }dV=0.  \label{ModE1}
\end{equation}
Thus, the explicit form of Einstein equation with power-law correction term
is obtained as
\begin{equation}
R_{\mu \nu }=8\pi G\left( T_{\mu \nu }-\frac{1}{2}Tg_{\mu \nu }\right)
\left( 1+\mathcal{M}_{\alpha }\right) ,  \label{ModE2}
\end{equation}
where
\begin{equation}
\mathcal{M}_{\alpha }=\left( \frac{\alpha -4}{\alpha -2}\right)
\frac{K_{\alpha }}{4\pi A^{\alpha /2-1}},  \label{ModE3}
\end{equation}
represents the correction term to Einstein equation. In the absence of
power-law correction ($\alpha =0=K_{\alpha }$), one recovers the standard
Einstein equation in general relativity. For large horizon areas, the
power-law correction becomes small and the standard Einstein equation is
recovered.

On the other side, if we take the logarithmic correction to entropy, the
function $s(A)$ becomes
\begin{equation}
s(A)=-\beta \ln {\frac{A}{4\ell _{p}^{2}}}+\gamma \frac{\ell _{p}^{2}}{A}+%
\mathrm{const}.  \label{SLog}
\end{equation}
In this case the right hand side of Eq. (\ref{ModE}) can be written as
\begin{equation}
-4\ell _{p}^{2}\int e^{\phi }\left( \frac{\partial s}{\partial A}\right)
\triangledown \phi .dA=-4\ell _{p}^{2}GM\left( \frac{\beta }{A}+\frac{\gamma
}{2}\frac{\ell _{p}^{2}}{A^{2}}\right).  \label{SLog2}
\end{equation}
It is matter of calculation to show that for the logarithmic entropy
correction, Eq. (\ref{ModE}) reduces to
\begin{equation}
\int \left[ R_{\mu \nu }-8\pi G\left( T_{\mu \nu }-\frac{1}{2}Tg_{\mu \nu
}\right) +8\ell _{p}^{2}G\left( \frac{\beta }{A}+\frac{\gamma }{2}\frac{\ell
_{p}^{2}}{A^{2}}\right) \left( T_{\mu \nu }-\frac{1}{2}Tg_{\mu \nu }\right) %
\right] N^{\mu }\xi ^{\nu }dV=0.  \label{ModE11}
\end{equation}
Therefore, we can find the modified Einstein equation as
\begin{equation}
R_{\mu \nu }=8\pi G\left( T_{\mu \nu }-\frac{1}{2}Tg_{\mu \nu }\right)
\left( 1-\Gamma\right) ,  \label{ModE22}
\end{equation}
where
\begin{equation}
\Gamma=\frac{\ell _{p}^{2}}{\pi }\left( \frac{\beta }{A}+\frac{\gamma }{2}%
\frac{\ell _{p}^{2}}{A^{2}}\right) .  \label{ModE33}
\end{equation}
We see that the Einstein equation will modified accordingly with the
logarithmic correction in the entropy-area expression. It is clear that
without correction terms ($\beta =\gamma =0$), Eq. (\ref{ModE22}) reduces to
the familiar Einstein equation. Again, for large horizon areas, $A$, the
correction becomes small and the standard Einstein equation is recovered.

\section{Conclusion\label{Sum}}

The entropy of a black hole is proportional to its horizon area
and obeys the well-known Bekenstein-Hawking area law
$S=A/4\ell_p^2$ \cite{HB}. However, the area law is a
semi-classical result and there is no reason to believe it to be
the complete answer conceivable from a correct quantum gravity
theory. Therefore, it is imperative for any approach to quantum
gravity to go beyond area law and provide generic subleading
corrections. Logarithmic corrections to area law, arises from loop
quantum gravity due to thermal equilibrium fluctuations and
quantum fluctuations. Another type of correction, usually called
power-law correction, appears in dealing with the entanglement of
quantum fields in and out the horizon \cite{Sau}. The entanglement
entropy of the ground state obeys the Bekenstein- Hawking area
law. However, a correction term proportional to a fractional power
of area results when the field is in a superposition of ground and
excited states. In other words, the excited state contributes to
the power-law correction, and more excitations produce more
deviation from the area law \cite{sau1}.

In this paper, we considered the general corrections to the area
law and studied the modification of several equations related to
gravity theory accordingly. We followed the Verlinde's viewpoint
and proposed that gravity is a kind of entropic force caused by
the changes in the information associated with the positions of
material bodies. Following the logic of \cite{Ver}, we derived the
correction terms to MOND theory of gravitation and Einstein field
equations. As two special cases we considered the logarithmic and
power-law corrections to entropy and found the explicit form of
the obtained modified equations. The obtained results in this
paper are quite general and can be applicable to any kind of
correction terms to entropy which may be find in the future.

%%%%%%%%%%%%%%%%%%%%%%%%%%%%%%%%%%%%%%%%%%%%%%%%%%%%%%%%%%%%%%%%%%%%%%%
\acknowledgments{We thank the referee for constructive comments.
This work has been supported by Research Institute for Astronomy
and Astrophysics of Maragha.}
%%%%%%%%%%%%%%%%%%%%%%%%%%%%%%%%%%%%%%%%%%%%%%%%%%%%%%%%%%%%%%%%%%%%%%%%%%%

\end{document}